# Hybrid Learning Aided Technology-Rich Instructional Tools - A Case Study: Community College of Qatar

Muhammad Jamal Shehab[1], Mosa Alokla[2], Mais Alkhateeb[3], and Mohammad Alokla[4]

[1]Electrical Engineering Department, Qatar University

[1]Information Technology Department, Community College of Qatar

[2]Teaching and Learning Center (TLC), Community College of Qatar

[3]Math and Science Department, Community College of Qatar

[4]Information Technology Department, Damascus University

## Abstract

Educational Institutions have an essential role in promoting the teaching and learning process, within universities, colleges, and communities. Due to the recent coronavirus COVID 19 pandemic, many educational institutions adopted hybrid learning (HL), which is a combination of classic and online learning. It integrates the advantages of both, and it is a fundamental factor to ensure continued learning. Technological innovations such as HL are changing the teaching process, and how students, lecturers, and administrators interact. Based on this, the Community College of Qatar (CCQ) focused on researching the structures and elements related to the adoption of HL. Thus, the goal of this research paper is to reveal the impact of HL on the learning process in CCQ, and the effective instructional technology (INST) tools required for a successful HL program. Our research questions for the survey were designed to measure the opinions of the students, instructors, and administrators about the HL program. It is observed from the results that the majority of students, instructors, and administrators showed a positive attitude toward HL, but some had negative views and experienced challenges. The results were analyzed and discussed to better utilize HL to meet the growing demands of the community.

# Introduction

The breakthrough out of the coronavirus COVID-19 pushed hybrid learning (HL) forward since it mingles the advantages of both face-to-face (F2F) classroom teaching and online learning (see Figure 1). HL is now considered a key element to ensure continued education according to research (UNESCO, 2020). The fundamental redesign of the educational process at the Community College of Qatar (CCQ) based on the HL program requires the need to address the challenges associated with this new program since the motif of HL is not only a simple mix of F2F and online instruction but rather a focus on enhancing the achievement of learning and teaching objectives by employing "appropriate" instructional technologies to match the "appropriate" learning to the "appropriate" person at the "appropriate" time according to research (Bonk & Graham, 2006). With the increasing advancements in instructional technology (INST), academic institutions over the world seek to integrate technologies and resources to facilitate the teaching and learning process, make it more practical, interesting, and attractive to enhance the active learning strategies (ALSs), especially in the presence of hybrid learning which combines both F2F and online instruction simultaneously.

In research from Community College of Qatar (CCQ) (2020), one of CCQ's goals is to aspire to motivate and transform to become an educational institution that provides students with the latest educational tools and platforms, and the adoption of HL accelerates the process CCQ et al. The idea may be intuitively simple, but practically it is complex. The reason is that the instructor needs to reduce the F2F meetings and replace it with a huge amount of instructional time in addition to online activities according to (Allan, 2006). Hybrid classes focus on online activities to assist students in learning and practicing, in addition to their participation in a F2F classroom. Thus, the students will be active, and this will strengthen their technological background by using the computer, smartphone, tablet, and learning management systems (LMS) such as Blackboard. HL provides students with flexibility, and this will assist students in studying faster and possessing worthy INST skills. Further, HL will furnish the students with the required skills to think and work independently. This will strengthen them to overcome the challenges in the future, which is aligned with the Qatar National Vision (QNV) 2030 according to Community College of Qatar (CCQ) (2020).

**Figure 1**

*Hybrid Learning*

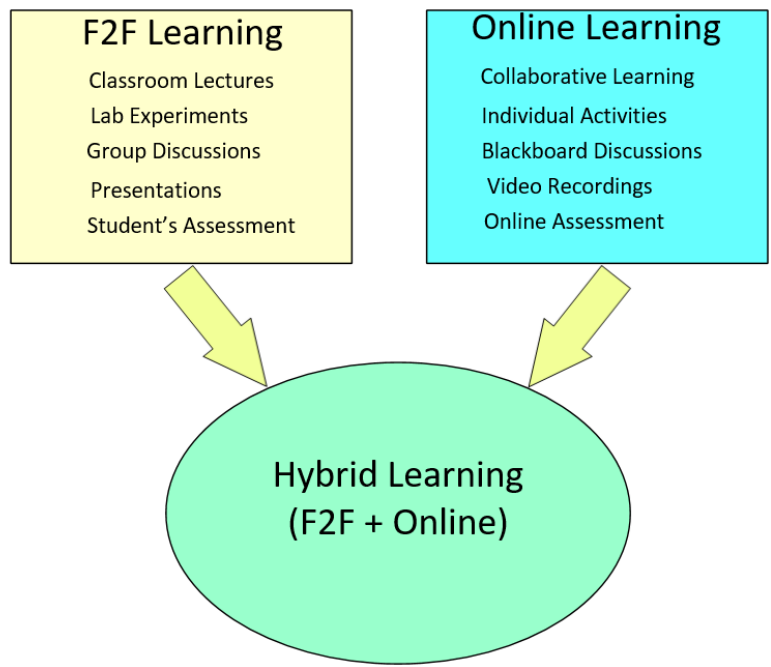

*Note*. Hybrid Learning mixes both F2F and online learning.

## Literature Review

The components of hybrid learning are F2F learning and online learning according to research (Yu, 2021). There is a difference between learning in classrooms F2F and online virtual sessions (Suwannaphisit et al., 2021). In F2F classroom learning, the communication is verbal with real-time body language, visual cues, and F2F context. However, in online virtual sessions, the communication between the instructor and the students occurs in virtual form with video conferencing handwritten text format without the presence of the body language and visual cues factors according to (Qiuyun, 2009). In research from Shetu et al. (2021), the authors proposed and developed an effective e-learning scheme to support the online learning system and facilitate self-learning for students. The proposed e-learning model enables learners to learn computer skills and upgrades interactive skills. Many other recent studies investigated the HL model during the COVID-19 pandemic. HL was involved in education sectors such as nursing (Leidl et al., 2020; Berga et al., 2021; Ainslie et al., 2021; Tebbs et al., 2021; Z. Yu et al., 2021) and medical as well (Vincent et al., 2021; Joos et al., 2022). In research from Garg et al. (2021), the virtual and online classrooms during the COVID-19 pandemic did not replace the hands-on experience through the F2F interaction with senior staff. To overcome this obstacle, they organized a hybrid workshop that involves simulation-based learning modules, with virtual and direct interaction with surgeons throughout the live surgeries. Also, didactic lectures were used to assist delegates in the perception of the nuances of neurosurgery. Furthermore, the authors (Lapitan et al., 2021) developed an effective HL strategy during the pandemic, which facilitated the transition from F2F learning to online learning, and it is based on five key elements denoted as Discover, Learn, Practice, Collaborate, and Access. Nonetheless, the advancements and deployments of instructional technologies in the learning process are evolving. Nowadays, students and instructors are familiar with digital technology. In research from Ifenthaler and Widanapathirana (2014), instructors are encouraged to utilize the INST tools and Information technology (IT) applications in teaching to encourage and stimulate students' learning. HL involves various learning strategies since it provides educational productivity, reduced cost, personal growth, knowledge awareness, joint participation, and collaboration. Further, HL resolves physical attendance problems and simplifies corrections according to (Mustapa & Ibrahim, 2015). It reinforces learning and adds motivation, interactivity, social interactions, and leads to enhanced feedback and use of study materials (Sun & Qiu, 2017). HL is anticipated to become the new learning model that utilizes various media resources to empower the interaction between the students (Graham & Woodfield, 2013). For example, it provides significant and motivating learning tools through various teaching methods such as social networking, blog, webinars, forums, live chats, etc. which provides students with more opportunities to express and give feedback (Graham, 2013). Learning management systems (LMS)s (Bervell et al., 2021) such as Moodle (Lebeaux et al., 2021), WebCT, and Blackboard facilitate the migration toward HL. INST tools are utilized to enable collaborative learning between the lecturers and the students (Jr. et al., 2017). HL program involves online activities (see Figure 1) such as collaborative learning, individual activities, studying video recordings, online presentations, Blackboard discussions, and online assessment (B. Anthony & A. Kamaludin, 2019). More, it includes F2F instruction such as lab experiments, class lectures, group discussions, presentations, and class assessments (Sun & Qiu, 2017). The recent development of HL focused on optimizing the outcome of the learning process. Therefore, previous studies evaluated the effectiveness of HL compared to online teaching and traditional teaching (van Laer & Elen, 2018). Moreover, while there are research studies on HL, studies that focused on the implementation and adoption of HL are still limited, and this is a gap to be addressed. Furthermore, some studies on HL focused on surveys for students only (Alabdulkarim, 2021; Mali & Lim, 2021). However, in our case study, we will focus on CCQ adoption to HL, how the INST tools support HL, and how students, lecturers, and administrators interact when HL is adopted. We performed three surveys for administrators, instructors, and students who are involved in the HL program, and based on these surveys we will discuss the cons and pre-cons of the HL and how to address the findings of the survey to better employ HL to meet the growing demands of the college and the community.

**Research Aim and Objectives**

This research aims to investigate the impact of HL in the educational process in CCQ, in addition to exploring the information and communication (ICT) technologies used to enhance the educational process and guarantee the success of the program. Our research will be based on the following questions:

• Is HL a suitable learning strategy that would best suit the students?

• What are the INST tools needed in the HL program at CCQ?

• Will the students meet the learning objectives if the HL program is adopted?

To answer these research questions in this study, a survey was designed and developed to evaluate the student's awareness and involvement in the HL program. The findings are then discussed and analyzed to set recommendations to empower the HL program in CCQ.

**Research Methodology**

This research paper is a case study approach for CCQ. A quantitative method was utilized in this paper to meet the objectives of this research. Thus, surveys are performed as an evaluation tool to measure the students' and instructors' level of awareness, concerns, opinions, behaviors, and their willingness to participate in the HL program. This is to understand to what extent the students and instructors are satisfied with this program, and data is generated based on these surveys. Moreover, the appropriate information is collected through research and literature review to answer the research questions as well. The literature review on HL was conducted to satisfy the requirement of this study. Finally, recommendations are given based on the findings to answer the research questions and to verify the effectiveness of the HL strategy, taking into consideration the limitations encountered during the analytical process.

## Survey Analysis

### Survey Context

Three surveys for students, instructors, and administrators took place at CCQ in Qatar. Data was gathered from several hybrid courses taught by CCQ instructors during spring 2021. There are different courses taught in a hybrid format such as the Business Computer Applications course (BCIS1305) and Computer Fundamentals (ITEC 1300). The focus of the BCIS course is on business applications of software including Microsoft Word, Excel, PowerPoint, Teams, Streams, and business-oriented utilization of the Internet. It is a 3credit hours course, introductory, and needs no prerequisites. This course is designed to equip students with learning environments to master the fundamentals of computer skills which are important in dealing with technology as an educational tool in their future studies. It was taught by different instructors during spring 2021. The second course which is computer fundamentals provides an overview of computers and computer literacy. This course is also 3 credit hours and focuses on the history of computing, personal computers, and computer components including hardware, input, output, system software, and applications.

### CCQ INST Tools Utilized In The Learning Process

The main INST Tools used in CCQ for online and hybrid education are Blackboard, Banner, Microsoft Team, and Cisco WebEx.

- **Blackboard**: It is an LMS that includes many important zones such as the:

- **Content Zone**: This includes the instructor's contact info, course material, learning module, class recordings, syllabus, rubrics, and announcements.

- **Assessment Zone:** This includes quizzes, tests, assignments, surveys, and a review of a graded paper.

- **Communication Zone**: This includes the discussion board, emails, chats, and class rosters.

- **Student's Zone:** Students have access to check their grades, make a personal homepage, calendar, and profile. Blackboard assists in making the learning process paperless. Thus, the instructor can share with students the syllabus, lectures, power points, and video recordings. More, teachers can utilize Blackboard to assess students by giving homework, quizzes, and exams. This is a significant tool in online and hybrid learning which helps in managing the learning process effectively Qiuyun Lin (2008 – 2009).

2) **Banner:** Banner is student information systems (SIS)which are considered vital in the learning process. It is used to schedule classes and classify them whether it is F2F, online, or hybrid. This system can be also used to track attendance, extract attendance records, summarize class lists, view class schedules, view assignment history, add midterm and final grades.

3) **Microsoft Teams:** This software is one of the most popular and important Microsoft office 365 tools that serve participants for online virtual classes with many features and easy functions such as audio call, video conference, whiteboard, and webchat. Further, one can organize his meeting calendar and form teams. MS teams can be linked to Blackboard in which the instructor can share the MS teams meeting link with all students on Blackboard. This online virtual link on Blackboard can be used by students to join the MS team's online session. Moreover, MS teams offer breakout classrooms to aid instructors in establishing meaningful connections via smaller and comfortable group settings. This feature enables students to be separated into groups to enhance and foster collaboration, facilitate discussions, and motivate participation. Hence, MS teams' application enhances experience and education at CCQ, and it is very important for hybrid and online learning (Microsoft, 2021).

4) **Cisco WebEx**: Cisco WebEx is a modern platform used for online virtual classes with rich options and features in addition to its high performance. It facilitates virtual learning and HL which enhances the collaboration and engagement of the instructor and students. Furthermore, it enables you to have education anytime and anywhere for connected classrooms, and thus it improves the students' outcomes since it mixes synchronous and asynchronous learning (Virtual Classroom, 2021).

**Participants**

The participants in this survey were the instructors, students, and administrators. The number of students that participated in this survey in spring 2021 is 116 from two-hybrid courses which are the BCIS course and the Computer Fundamentals course. The BCIS hybrid course is divided into three classes with 46 students in the first class, 36 students in the second class, and 16 students in the third class. Further, the number of students in the Computer Fundamentals course is 18. Thus, the total number of students that participated in this survey is 116. These students are undergraduates from different disciplines and various departments. They had the experience of taking hybrid courses. Eighty-eight of the students are female and twenty-eight of the students are male, which means that the female students' percentage is 76 percent and the male students 'percentage is 24 percent. Their ages range between 20 and 50years old.  Further, the students' GPAs are between 2 and 4, with a median of 2.7 on a scale of 4. Moreover, the number of instructors and administrators that taught hybrid courses and participated in this survey is 18 and 23 respectively.  These instructors had the experience of teaching hybrid, online andF2F courses. Therefore, the total number of participants is 157 (see Figure 2).

**Figure 2**

*Total Number Of Participants In The Survey*

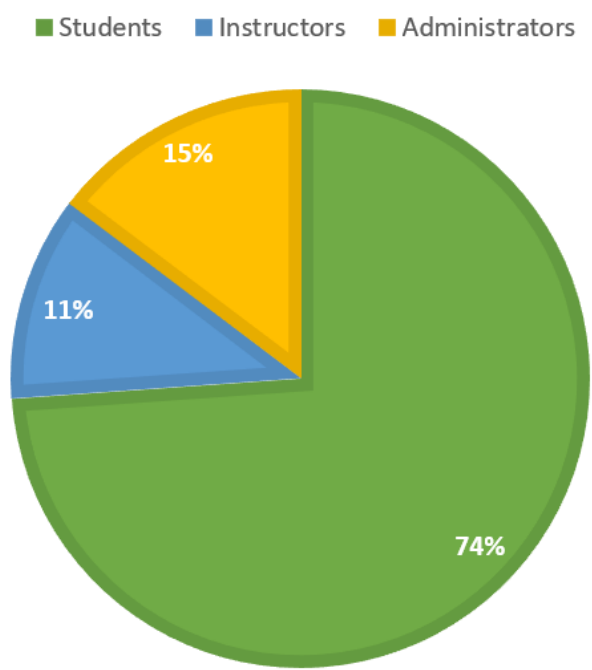

*Note*. The total number of participants in the survey is shown for students, instructors, and administrators.

**Information Collection and Analysis**

The information collected from CCQ staff was based on three hybrid course surveys for the instructors, students, and administrators.  These surveys were focused on three main sections using a 5-point Likert scale, the first section was about the educational technology background of the students and instructors as well as the benefit and usage of Blackboard tools for them. The second section was about the experience of students, instructors, and administrators with the CCQ HL program. Further, we have taken into consideration their recommendation for the HL program in the future. Data analysis included a quantitative method with descriptive statistics needed to analyze the survey questions. This included calculating the frequency, percentages, means, and standard deviations.

## Results And Discussions

The total number of participants in this survey is 157, around 74% of them are students, 15% are administrators, and 11 % are instructors (see Figure 2). Further, the percentage of females that participated in this survey is 62% whereas that of males is 38% (see Figure 3). The focus of these surveys was on the views of students, instructors, and administrators about the HL program in CCQ, and whether they found this program successful or not. Based on the survey results, the students, instructors, and administrators showed a positive attitude toward the HL process. We will start with the results of the first survey which was about the educational and informational technology background of the students and instructors in CCQ.

**Figure 3**

*Percentage of Males and Females participated in the Survey*

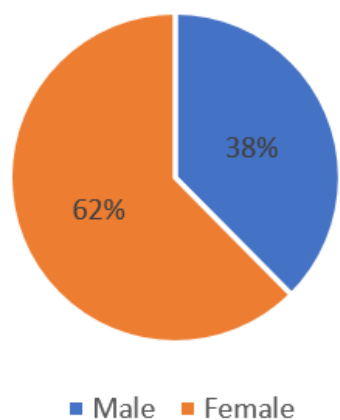

Percentage of Males and Females

*Note*. The percentage of males and females who participated in this survey is shown for each gender.

## First Survey about the Technology Background for Students and Instructors

Table 1 below shows descriptive statistics for this survey, and from these results, we can infer that the majority of the students and instructors have acceptable background InTechnology. The mean of the survey results in Table 1 for the student's educational technology background ranges between3.57 and 3.72 which is acceptable and the standard deviation ranges between 0.8 and 0.97, whereas for the instructors 'educational technology background the mean ranges between4 and 4.55 and the standard deviation ranges between 0.67and 0.89. This reveals that the instructors' informational technology background is solid. For the second part of the table Which is about the benefits of blackboard tools for the students and instructors, the mean ranges between 4.18 and 4.38 which reflects a very good indication for the students' interest in blackboard tools and that for instructors is between 4.17 and4.67. This shows the enormous benefit blackboard is offering as an INST tool.

**Table 1**

*Descriptive Statistics Of The Surveys Performed For The Hybrid Learning Program*

| Survey Items | Students (BCIS and Computer Fundamentals) N = 116 | | Hybrid Learning Instructors N = 18 | |
|---|---|---|---|---|
| **Educational Technology Background** | Mean (SD) | Percentage | Mean (SD) | Percentage |
| Microsoft Windows | 3.57 (0.8) | 54.31 % | 4.11 (0.87) | 66.66 % |
| Microsoft Word | 3.7 (0.9) | 62 % | 4.55 (0.68) | 88.88 % |
| Microsoft Excel | 3.72 (0.93) | 60.34 % | 4.17 (0.89) | 66.66 % |
| Microsoft Power Point | 3.60 (0.94) | 55.17 % | 4 (0.82) | 66.66 % |
| Microsoft Teams | 3.37 (0.94) | 43.10 % | 4.11 (0.74) | 77.77 % |
| Microsoft Streams | 3.28 (0.92) | 44.82 % | 4.39 (0.75) | 83.33 % |
| Internet | 3.49 (0.94) | 49.13 % | 4 (0.67) | 77.77 % |
| Email | 3.5 (0.97) | 50 % | 4 (0.82) | 66.66 % |
| Blackboard | 3.45 (0.95) | 47.41 % | 4.17 (0.69) | 83.33 % |
| Banner | 3.44 (0.95) | 46.55 % | 4.17 (0.83) | 72.22 % |
| Comfort Level with Technology | 3.40 (0.94) | 44.82 % | 4.33 (0.67) | 88.88 % |
| **Benefits of Blackboard Tools** | | | | |
| Announcements | 4.18 (0.77) | 77.58 % | 4.39 (0.75) | 83.33 % |
| Online Virtual Link | 4.29 (0.71) | 90.51 % | 4.33 (0.74) | 83.33 % |
| Course Syllabus | 4.33 (0.67) | 90.24 % | 4.33 (0.74) | 83.33 % |
| Course Documents | 4.31 (0.78) | 90.51 % | 4.22 (0.85) | 72.22 % |
| Assignments | 4.27 (0.80) | 88 % | 4.39 (0.68) | 83.88 % |
| Exams | 4.27 (0.79) | 88.8 % | 4.61 (0.59) | 94.44 % |
| Simulation Tests | 4.27 (0.81) | 84.48 % | 4.17 (0.83) | 72.22 % |
| Discussion Board | 4.28(0.8(1) | 86.20 % | 4.33 (0.47) | 100 % |
| Video Recordings | 4.29 (0.79) | 84.48 % | 4.67 (0.57) | 94.44 % |
| My Grades | 4.30 (0.71) | 85.34 % | 4.28 (0.56) | 94.44 % |
| Resources/Links | 4.31 (0.63) | 90.51 % | 4.28 (0.73) | 83.33 % |
| Calendar | 4.33 (0.72) | 85.34 % | 4.33 (0.63) | 88.88 % |
| Building Learning Module | 4.38 (0.66) | 89.65 % | 4.5 (0.76) | 83.33 % |

*Note.* These results reveal the percentages of instructors and students who selected strongly agree or agree on a 5-point Likert scale.

Besides, from the Microsoft Office Suite, we realize that the students are more experienced in word, excel, and PowerPoint more than teams and streams. The reason is that they used to utilize word, excel, and PowerPoint in their practical work or study, but they recently started using teams and streams for educational purposes in online and hybrid learning processes. Teams' software is used in online and hybrid learning for lectures, meetings, presentations, and exams. Further, stream software is used for the recordings of lectures where the instructor can upload the lecture recorded and the students can watch these recordings for lecture review.

The experience of students in Microsoft Office, internet, email, Blackboard, and Banner ranges between 43 % and62 %. The percentages in the table are calculated based on the students and instructors who selected" Strongly Agree "and" Agree" on a 5-point Likert scale. The highest percentage was for word, excel, and PowerPoint, and this is expected since the word is almost used by students and instructors daily for writing reports, letters, CVs, etc. More, PowerPoint is utilized by both instructors and students for delivering presentations, and excel is used for setting up charts, worksheets, and tables. It is for these reasons students are familiar with the word, excels, and PowerPoint more than teams and streams. Further, Table 1 reveals the opinion of the students and instructors toward the benefits of the Blackboard tools such as the announcements, online virtual link which is dedicated for the students to log in to the online lecture via Teams or WebEx, course syllabus, course documents, assignments, exams, simulation tests, discussion board, video recordings, grade center, resources/links, calendar, and building learning module. Results show that Blackboard tools are beneficial for both instructors and students since the majority of them indicated that Blackboard tools are beneficial for them. The percentages of the Blackboard tools benefits for students and instructors are between 72.22 % and 100 %. More, it is realized that all the instructors 100 % mentioned that the discussion board is beneficial for them. The reason is that they can explore the opinions of their students and enables them to cooperate and collaborate which makes the course more interesting and adds motivation. With a discussion board, shy students who do not participate in F2F discussions will be able to share their opinions with others online.

**Second Survey about the Hybrid Learning Experience for Students**

Table 2 shows descriptive statistics of the survey performed for the students about The hybrid learning program. In the table, some items show a positive impact of HL, and some items show a negative impact of HL. Similar to Table 1, the percentages of these items are calculated based on the students and instructors who selected" Strongly Agree" and" Agree" on a 5-point Likert scale.

We can realize from the results that the majority of the students 82.7% were satisfied with the existing HL model at CCQ, 81.9 % can control the pace of their learning, and 84.48 % believed that the online assignments are useful for them to understand the course and that the link between what they study online, and in-class was clear. Additionally, 89.19 % did not have any difficulty managing their time for the online part of the course, and more than three-quarters of them are encouraged to take a course that involves Blackboard. All these items show the positive effect of the HL program. On the other hand, some items reveal the negative influence of the HL program such as if the students were finding difficulty in following up with the online course, or if they find that the Blackboard discussion is useless for their online learning, or if they were unable to share ideas with other students online, or if they are spending a lot of time online more than F2F class. In our survey, the percentage for students that faced difficulty in the online course is 11.10 %, the students that faced problems in Blackboard discussions represent 6.11 %, the students that were unable to share ideas with other students represent 3.4 % and those who prefer to spend the time in class rather than spending too much time online represents 7.15 %. These percentages are very low, and they can be manageable. Thus, the instructional technology department along with instructors can train and educate these students to overcome these issues, especially since these percentages are very low for an HL program. In summary, this survey for students revealed that 76.72 % of the students support the HL process.

**Table 2**

*Descriptive Statistics Of The Survey Performed For The Students About The Hybrid Learning Program*

| Survey Items | Number of Students N = 116 |
|---|---|
| **Experience with HL Program** | |
| **Items showing the positive impact of HL** | |
| How satisfied are you with the current Hybrid learning model at CCQ? | 82.7 % |
| I can control the pace of my learning | 81.9 % |
| Online assignments helped me understand the course content | 84.48 % |
| The connection between what I did online, and in-class was clear | 84.48 % |
| I Did not have any difficulty managing my time for the online part of the course | 89.19 % |
| I would take another course that incorporates Blackboard | 75.86 % |
| **Items showing the negative impact of HL** | |
| The online course materials were difficult to follow | 11.10 % |
| I found participating in the online discussion board useless for my learning. | 6.11 % |
| I was unable to share ideas with other students regularly. | 3.40 % |
| The time I spent online would better have been spent in class. | 7.15 % |
| **Recommendation** | |
| I would recommend taking hybrid courses to other students. | 76.72 % |

*Note.* These results reveal the percentages of students who selected strongly agree or agree on a 5-point Likert scale.

**Third Survey about the Hybrid Learning Experience for Instructors**

The results of the admin staff survey reveal and acceptable results toward the HL program as shown in Table 3.  These staffs are neither instructors, nor students but they participated in the HL survey, some of them have roles such as higher management, IT team, security, and safety, etc.  Around 56.5 % were satisfied with the HL program and 77.77% of them recommended the program, 39.13 % were satisfied with the time students are spending at CCQ. Furthermore, the confidence percentage for the admin staff that the teachers can provide effective instruction in the HL model, motivating the students and assisting those who need help were 47.9 %, 56,5 %, and 43.4% respectively. Moreover, around 52.17 % expected that the HL model has a positive impact on the students' social, emotional, and well-being.  Add to this, 17.4 % of the admin staff observe that it is easy for them to support their families, friends, and loved ones in the HL model.  However, the same percentage of them believe that there are challenges in the HL model and they would like to see it improved.  Further, 52.17 % of the administrators are convinced of the changes happening in the program. Finally, 47.8 % of the admin staff recommended the HL program.

**Table 3**

*Descriptive Statistics Of The Survey Performed For The Instructors About The Hybrid Learning Program*

| Survey Items | Number of Instructors N = 18 |
|---|---|
| **Experience with HL Program** | Percentage |
| **Items showing the positive impact of HL** | |
| How satisfied are you with the current Hybrid learning model at CCQ? | 100 % |
| I would recommend other teachers to give hybrid courses. | 88.88 % |
| How confident are you that you can provide effective instruction? in the current learning model? | 77.77 % |
| How confident are you that you can motivate your students to learn in the current model? | |
| How confident are you that you can help your students who need it? the most academic support in the current learning model? | 88.88% |
| How easy is it for you to use the distance learning tools (video calls, learning applications, etc.)? | 100% |
| How easy is it to support other people in your life (family, friends, loved ones, etc.) with the current learning model? | 61.11% |
| What kind of effect is the current learning model having on your social-emotional well-being? | 55.55% |
| Is the current learning model working well and you would like to see continued? | 66.66% |
| In the past week, how many of your students regularly participated in your virtual classes? | 87.5 % |
| How about your students 'performance in the current learning model? | 83.12% |
| In the past week, how engaged have students been in your F2F classes? | 91.66 % |
| Items showing the negative impact of HL | |
| I found participation in the online discussion board useless for the students. | 4.03% |
| The time I spent online would better have been spent in class. | 3.02% |
| I was unable to share ideas with the students regularly. | 8.12% |
| Are there any challenges about the current learning model that you would like to see improved? | 7% |
| **Recommendation** | |
| I would recommend other instructors to give hybrid learning courses. | 88.17 % |

*Note.* These results reveal the percentages of instructors who selected strongly agree or agree on a 5-point Likert scale.

**Fourth Survey about the Hybrid Learning Experience for Administrators:**

The results of the admin staff survey reveal and acceptable results toward the HL program as shown in Table 4.  These staffs are neither instructors, nor students but they participated in the HL survey, some of them have roles such as higher management, IT team, security, and safety, etc.  Around 56.5 % were satisfied with the HL program and 77.77% of them recommended the program, 39.13 % were satisfied with the time students are spending at CCQ. Furthermore, the confidence percentage for the admin staff that the teachers can provide effective instruction in the HL model, motivating the students and assisting those who need help were 47.9 %, 56,5 %, and 43.4% respectively. Moreover, around 52.17 % expected that the HL model has a positive impact on the students' social, emotional, and well-being.  Add to this, 17.4 % of the admin staff observe that it is easy for them to support their families, friends, and loved ones in the HL model.  However, the same percentage of them believe that there are challenges in the HL model and they would like to see it improved.  Further, 52.17 % of the administrators are convinced of the changes happening in the program. Finally, 47.8 % of the admin staff recommended the HL program.

**Table 4**

*Descriptive Statistics Of The Survey Performed For The Administrators About The Hybrid*

*Learning Program*

| Survey Items | Number of Administrators N = 23 |
|---|---|
| Experience with HL Program | Percentage |
| How satisfied are you with the current Hybrid learning model at CCQ? | 56.5 % |
| How satisfied are you with the time students are currently spending for learning personally at CCQ? | 39.13 % |
| How confident are you that teachers can provide effective instruction in the current learning model? | 47.8 % |
| How confident are you that teachers can motivate students to? learn in the current model? | 56.5 % |
| How confident are you that teachers can help students who need the most academic support in the current learning model? | 43.4 % |
| Do you expect that the current learning model will have a positive effect on the student's social-emotional wellbeing? | 52.17 % |
| How easy is it to support other people in your life (family, friends, loved ones, etc.) with the current learning model? | 17.4% |
| How satisfied are you with the current learning model? Do you wish to continue? | 52.17 % |
| Are there any challenges about the current learning model that you would like to see improved? | 17.4 % |
| How convinced are you of the changes occurring in the Hybrid? learning program? | 52.17 % |
| Recommendation | |
| I would recommend to my friends to take hybrid courses | 77.77% |

*Note.* These results reveal the percentages of administrators who selected strongly agree or agree on a 5-point Likert scale.

## Survey Findings and Author's Point of View

When viewing the statistical results of the four surveys, we can conclude that the findings are encouraging, since the majority of the participants recommended the HL program, and they were satisfied with the HL model. This implies that the HL strategy is suitable for the students, especially when dealing with the COVID19 situation. Further, the INST tools utilized in the HL process are very important and play a significant role in the success of the program. Based on this, we make sure that we possess and utilize the latest technology in the education process.

### Discussions

From the survey results, we can view that the majority of the students are meeting the learning objectives in the HL program, and those who are not meeting the learning objectives have a low percentage and are being assisted by the instructors and the instructional technology team. The majority of the students and instructors are familiar with educational technology as we can infer from Table 1. Table 2, Table 3, and Table 4 reveal that the majority of the students, instructors, and admins have a good experience in the HL program since the items showing positive impact have high percentages and the items showing negative impact have low percentages. We can also infer from the results that the majority of students, instructors, and admins recommended the HL with percentage of 76.72%, 88.17 %, and 77.77% respectively. These percentages are considered very high and encouraging, since this program was new to students, and they were not used to it. The reason for this is that the selected participants have a good informational and educational background, and this assisted them to use the technology in their study without feeling that there is so much difference between F2F and online. Whereas, if we selected participants from another college who do not have a strong technological background or if they have a strong educational background, but their college did not offer for them the required instructional and technological tools, the results would have been different.

From our point of view, the effective HL environment is significant in guaranteeing innovative educational approaches via the adoption of the latest technology in learning and teaching. Many key elements aid in building effective and flexible HL environments such as the inspection of the student's learning outcomes, educational and IT background.

In planning the design, adoption, and implementation of the HL program we are attentive to the implications and indications raised by this survey which is a directional guidance for the design, adoption, and implementation of the Program. Universities, colleges, and schools in Qatar need to be heedful of the interconnection among the design features, student characteristics, and learning outcomes/results which are clear indicators of the effectiveness of the HL program. Artificial Intelligence (AI) can play a significant role in the success of the HL program. The reason is that it facilitates Asynchronous Learning Activities (ASA) that are suitable for the student's needs and does not demand constant supervision from the instructors. It assists the learners to get back on track after the summer vacation. Thus, it saves the instructors effort and time which makes the HL process more manageable for them. This assists students in getting individualized problems that focus on the skills that they have difficulty or challenges with it. ASA includes recorded lectures/presentations, discussion boards, videos, slideshows, social media groups, cloud collaborative documents, and email, etc. In other words, students can practice in a safe environment without being penalized or losing marks for making mistakes.

Thus, instructors can rapidly determine which students are progressing and which students are suffering and need help. This reveals the importance of AI in the HL process. Furthermore, increasing the quality of HL will contribute to achieving the Sustainable Development Goals (SDGs). The reason is that enhancing the quality of hybrid education will raise the number of qualified instructors to teach and educate the students. Thus, optimizing the teaching and education quality (SDG 4 – quality education) will help in achieving other objectives such as increasing the high-skilled workforce staff. This will result in productive employment, which in turn reinforces the economic growth (SDG 8 – economic growth) and build powerful and robust institutions (SDG16 – strong growth). In summary, fulfilling the SDGs in hybrid education is challenging, but with reliable findings, a solid plan, optimal design for a learning environment, and top-quality teachers for the HL program this challenge can be controlled.

## Conclusion And Future Recommendations

This research study on HL offers valuable insight regarding research related to HL program in Qatar. However, this study constructed the factors that affect the students, instructors, and administrators toward adopting the HL program.  Thus, a questionnaire survey was employed as a research method for data collection.  The surveys were carried out in CCQ, and they identified the factors that affect the perception of administrators, students, and instructors' readiness towards HL adoption.  These factors can be utilized to develop and build a model to examine instructors, students, and administrators simultaneously towards HL adoption and implementation in Qatar and countries abroad. Nonetheless, this is a small-scale research study that involves 157 participants from the CCQ HL program. Thus, the findings in this research paper cannot be generalized, especially since all the participants were from the same college and the process at CCQ is still in its infancy. Therefore, our future work is anticipated to be conducted on a large scale including multiple colleges, schools, and universities inside and outside Qatar to investigate the opinion of students about the program. Currently, in our study, we included factors and elements related to the informational technology background of the students and instructors, as well as their experience in the HL program. More, we involved a sample from the admin staff to participate in our survey. However, in our future study, we will make sure to explore each item mentioned in this survey in detail and include other new items related to the process. Furthermore, in addition to the surveys, we will include case studies and F2F interviews with students, in-depth literature reviews about hybrid and blended learning, experimental methods which involve learning management system (LMS)dataset, and we will discuss the importance of AI, machine learning (ML), and Sustainability in the HL process. Therefore, it is anticipated that our large-scale research study would generate reliable findings in the area of HL education. In our point of view, in future research, we need to fully include plentiful determinants of learning and teaching hybrid courses.